\documentclass[aps,prd,preprint,showpacs,showkeys]{revtex4}
\usepackage{graphics} 
\newcommand{\e}{\mathrm{e}}
\newcommand{\ep}{\epsilon}
\newcommand{\vev}[1]{\left\langle #1 \right\rangle}
\newcommand{\fmslash}[1]{\hbox{$#1$\kern-0.5em\raise0.3ex\hbox{/}}}
\begin{document}
\preprint{KOBE-TH-04-03}  
\title{Two dimensional non-linear sigma
models\\ as a limit of the linear sigma models} 
\author{Hidenori Sonoda}
\affiliation{Physics Department, Kobe University, Kobe 657-8501, Japan}
\email[E-mail address: ]{sonoda@phys.sci.kobe-u.ac.jp} 
\date{August 2004, revised December 2004}
\begin{abstract}
We show how to obtain the $O(N)$ non-linear sigma model in two
dimensions as a strong coupling limit of the corresponding linear
sigma model.  In taking the strong coupling limit, the squared mass
parameter must be given a specific coupling dependence that assures
the finiteness of the physical mass scale.  The relation discussed in
this paper, which applies to the renormalized theories as opposed to
the regularized theories, is an example of a general relation between
the linear and non-linear models in two and three dimensions.
\end{abstract}
\maketitle

The purpose of this short paper is to elucidate the relation between
the $O(N)$ linear and non-linear sigma models in two dimensions.  We
are especially interested in obtaining an exact formula that gives the
\textbf{renormalized} non-linear models as a strong coupling limit of
the \textbf{renormalized} linear models.

The relation between the linear and non-linear sigma models is well
known in the case of three and four dimensions.  In four dimensions,
the two models are equivalent\cite{hasenfratz, z-j, soto, s99} up to
differences suppressed by inverse powers of a momentum
cutoff.\footnote{The first three papers show the equivalence of a
four-Fermi theory with a Yukawa theory.  The reason for this
equivalence is the same as for that between the $O(N)$ linear and
non-linear sigma models.}  In three dimensions, the renormalized
non-linear model with one parameter is obtained as a strong coupling
limit of the renormalized linear model with two parameters.  This
relation is analogous to the relation we will find for the two
dimensional theories, and for the reader's convenience we will review
the well-known relation in appendix A.

The main difference of the two dimensional theories from the three and
four dimensional theories is the lack of symmetry breaking for $N \ge
3$.  For the latter theories, a critical point plays an important role
for the relation between the linear and non-linear models, but in the
two dimensional case, there is no critical point.

To obtain an exact formula that relates the \textbf{renormalized}
linear model to the \textbf{renormalized} non-linear model, we first
review the relation for the \textbf{regularized} theories.  For
regularization we use a $D$-dimensional cubic lattice.  We will not
specify the dimension of the lattice till later.

The lattice action for the $O(N \ge 3)$ linear sigma model is given by
\footnote{The Boltzmann weight is given by $\e^S$.}
\begin{equation}
S = - \sum_{\vec{n}} \left[
 \frac{1}{2} \sum_{i=1}^D \left( \phi^I_{\vec{n}+\hat{i}} -
 \phi^I_{\vec{n}}
 \right)^2 + m_0^2 \frac{\left( \phi^I_{\vec{n}} \right)^2}{2}
 + \frac{4 \pi \lambda_0}{N-2}
 \frac{\left(\left(\phi^I_{\vec{n}}\right)^2\right)^2}{8}
 \right]
\end{equation}
where the vector $\vec{n}$ consists of $D$ integral coordinates of a
lattice site.  The vector $\hat{i}$ is the unit vector in the $i$-th
direction.  The field $\phi^I_{\vec{n}}$ at each lattice site is an
$N$-dimensional real vector with no constraint on its range.  We
suppress the summation symbol over a repeated index $I = 1, \cdots, N$.

Assuming $m_0^2 < 0$, we introduce
\begin{equation}
v_0^2 \equiv \frac{- 2 m_0^2}{\frac{4\pi \lambda_0}{N-2}} > 0
\end{equation}
in terms of which we can rewrite the action as
\begin{equation}
S = - \frac{1}{2} \sum_{\vec{n}}
 \left[ \sum_{i=1}^D \left( \phi^I_{\vec{n}+\hat{i}}
 - \phi^I_{\vec{n}} \right)^2 + \frac{\pi \lambda_0}{N-2}
  \left( \left(\phi^I_{\vec{n}}\right)^2 - v_0^2 \right)^2 \right]
\end{equation}
ignoring an inessential additive constant.  We rescale the field as
\begin{equation}
\phi^I_{\vec{n}} \longrightarrow v_0 \phi^I_{\vec{n}}
 \label{rescale}
\end{equation}
so that
\begin{equation}
S = - \frac{v_0^2}{2} \sum_{\vec{n}}
 \left[ \sum_{i=1}^D \left( \phi^I_{\vec{n}+\hat{i}}
 - \phi^I_{\vec{n}} \right)^2 + \frac{\pi \lambda_0 v_0^2}{N-2}
 \left( \left(\phi^I_{\vec{n}}\right)^2 - 1 \right)^2 \right]
\end{equation}

There are two ways of obtaining a non-linear sigma model:
\begin{enumerate}
\item $\lambda_0 \to \infty$ limit\\Using
\begin{equation}
\lim_{\lambda \to \infty} \sqrt{\frac{\lambda}{\pi}} \,\e^{- \lambda x^2}
= \delta (x)
\end{equation}
	we obtain the non-linear sigma model with the action
\begin{equation}
S = - \frac{N-2}{4 \pi g_0} \sum_{\vec{n}} \sum_{i=1}^D
 \left( \Phi^I_{\vec{n}+\hat{i}} -
 \Phi^I_{\vec{n}} \right)^2 \label{nonlinear}
\end{equation}
	where $\Phi^I_{\vec{n}}$ is a unit vector, and 
\begin{equation}
\frac{1}{g_0} = \frac{4\pi}{N-2} \frac{v_0^2}{2} = \frac{- m_0^2}{\lambda_0}
\label{latticegzero}
\end{equation}
\item \label{second}$v_0 \to \infty$ limit\\ In this limit, the
squared mass $m_0^2$ goes to $- \infty$.  We obtain (\ref{nonlinear}),
but only for the limiting case of $g_0 \to 0$.
\end{enumerate}

The first limit is what we usually have in mind as the relation
between the linear and non-linear sigma models.  We find that keeping
the ratio $\frac{m_0^2}{\lambda_0}$ fixed, the non-linear sigma model
is obtained in the strong coupling limit $\lambda_0 \to \infty$.  This
result is fine as long as we are interested only in regularized
theories.  But if we are interested in the relation between the
renormalized theories, the second limit is more useful, because the
continuum limit of the non-linear sigma model in two dimensions calls
for the limit $g_0 \to 0$ due to asymptotic freedom.

This is the right moment to recall how to obtain the continuum limit
of the two-dimensional non-linear sigma model.  Consider the two-point
function as an example.  Using the lattice theory (\ref{nonlinear})
for $D=2$, the continuum limit of the two-point correlation function
is given by\footnote{Here $\vec{r}$ is an arbitrary vector, hence
$\vec{r} \e^t$ is not necessarily integral.  If $t$ is big enough, we
can always approximate $\vec{r} \e^t$ by an integral vector.}
\begin{equation}
\vev{ \Phi^I (\vec{r}) \Phi^J (\vec{0}) }_g
 \equiv 
 \left(\frac{g}{1 + c g}\right)^{2 \gamma} \lim_{t \to \infty}
 t^{2 \gamma} \vev{\Phi^I_{\vec{n}=\vec{r}\e^t}
 \Phi^J_{\vec{0}}}_{g_0(t)}
 \label{limit}
\end{equation}
On the right-hand side, the parameter $g_0 (t)$ is given a specific
dependence on the logarithmic scale parameter $t$:
\begin{equation}
\frac{1}{g_0 (t)} \equiv t + c \ln t - \ln \Lambda (g)
\end{equation}
where 
\begin{equation}
c \equiv \frac{1}{N-2}
\end{equation}
and the mass scale $\Lambda (g)$ is defined by
\begin{equation}
\Lambda (g) \equiv \e^{- \frac{1}{g}}
 \left( \frac{g}{1 + c g} \right)^{-c}
\label{Lambda}
\end{equation}
The constant $\gamma$, which is the coefficient of the 1-loop
anomalous dimension, is given by
\begin{equation}
\gamma \equiv \frac{N-1}{2(N-2)}
\end{equation}
It is easy to check that the continuum limit satisfies the
renormalization group (RG) equation
\begin{equation}
\vev{\Phi^I (\vec{r} \e^{-\Delta t}) \Phi^J (\vec{0})}_{g +
\Delta t (g^2 + c g^3)} = \Delta t \cdot 2 \gamma g
 \vev{\Phi^I (\vec{r} \e^{-\Delta t}) \Phi^J (\vec{0})}_g
\label{nonlinearRG}
\end{equation}
where $\Delta t$ is infinitesimal.\footnote{Note that the renormalized
coupling constant $g$ and renormalized field $\Phi^I$ are chosen such
that the two-loop beta function $g^2+c g^3$ and the one-loop anomalous
dimension $\gamma g$ become exact.}

For the two-point function on the right-hand side of
Eq.~(\ref{limit}), we can replace the $g_0 \to 0$ limit of the
non-linear sigma model by the $v_0 \to \infty$ limit of the linear
sigma model.  Since only the long distance limit of the linear sigma
model is necessary, we might as well replace the linear sigma model on
the lattice by its continuum limit.

Let us note that the continuum limit of the linear sigma model is
parametrized by the squared mass $m^2$ and the self-coupling constant
$\lambda$.  The two-point function $\vev{\phi^I (\vec{r}) \phi^J
(\vec{0})}_{m^2,\,\lambda}$ satisfies the RG equation
\begin{equation}
\vev{\phi^I (\vec{r} \e^{-\Delta t}) \phi^J (\vec{0})}_{\e^{2\Delta t}
 m^2 + \Delta t \, C \lambda,\, \e^{2 \Delta t} \lambda}
 = \vev{\phi^I (\vec{r}) \phi^J (\vec{0})}_{m^2,\, \lambda}
 \label{linearRG}
\end{equation}
where
\begin{equation}
C = \frac{N+2}{N-2} \label{C}
\end{equation}

We now replace the right-hand side of (\ref{limit}) by the two-point
function of the renormalized linear sigma model as
\begin{equation}
\vev{\Phi^I_{\vec{n}=\vec{r}\e^t} \Phi^J_{\vec{0}}}_{g_0 (t)}
\longrightarrow \frac{z}{t} \vev{\phi^I (\vec{r}\e^t)
\phi^J (\vec{0})}_{m^2 (t,\lambda; g),\, \lambda}
\end{equation}
where $z$ is a normalization constant, and the inverse power of $t$ is
due to the change of normalization (\ref{rescale}) and $v_0^2 \propto
t$.  Hence, we obtain the following relation
\begin{equation}
\vev{\Phi^I (\vec{r}) \Phi^J (\vec{0})}_g
 = z \left(\frac{g}{1+cg}\right)^{2 \gamma}
 \lim_{t \to \infty} t^{2 \gamma - 1} \vev{\phi^I (\vec{r} \e^t)
 \phi^J (\vec{0})}_{m^2 (t, \lambda; g),\, \lambda}
\label{relation}
\end{equation}
For a given $\lambda$, $m^2 (t, \lambda; g)$ is given by
\begin{equation}
\frac{- m^2 (t, \lambda; g)}{\lambda} \equiv t + c \ln t - \ln \Lambda (g) -
 \frac{C-1}{2} \, \ln \lambda
 \label{m2tdep}
\end{equation}
where we have shifted $t$ by a finite amount proportional to $\ln
\lambda$.  As will be explained shortly, this shift is necessary to
make the right-hand side of (\ref{relation}) independent of the
arbitrary choice of $\lambda$.

To summarize so far, the ``derivation'' of the relation
(\ref{relation}) consists of three ingredients:
\begin{enumerate}
\item In the $v_0 \to \infty$ limit, the linear sigma model on the
lattice gives the $g_0 \to 0$ limit of the non-linear sigma model on
the lattice.
\item The continuum limit of the non-linear sigma model can be
constructed in the limit $g_0 \to 0$.
\item The long distance limit of the linear sigma model on the lattice
can be replaced by the long distance limit of the renormalized linear
sigma model.
\end{enumerate}

Though the validity of each ingredient seems sound, it must be
admitted that the derivation of the relation (\ref{relation}) is not
sufficiently rigorous.  To augment the rigor of the derivation, we
make the following two consistency checks:
\begin{enumerate}
\item \textbf{RG equation for the non-linear sigma model}: Applying the
RG equation (\ref{linearRG}) of the linear sigma model to the
right-hand side of (\ref{relation}), we can derive
the correct RG equation (\ref{nonlinearRG}).
\item \textbf{No dependence on the choice of $\lambda$}: The left-hand
side of (\ref{relation}) has no $\lambda$ dependence.  Hence, the
right-hand side should be independent of $\lambda$.  We verify this
independence in the following.

Under the infinitesimal change from $\lambda$ to $\lambda \e^{2 \Delta
t}$, we find from (\ref{m2tdep})
\begin{equation}
m^2 (t, \lambda \e^{2 \Delta t}; g)
= \e^{2 \Delta t} m^2 (t + \Delta t, \lambda; g) + \Delta t \, C \lambda
\end{equation}
Hence, we find 
\begin{eqnarray}
 &&\lim_{t \to \infty} t^{2 \gamma - 1} \vev{ \phi^I (\vec{r} \e^t)
 \phi^J (\vec{0})}_{m^2 (t, \lambda \e^{2 \Delta t}; g),\,
 \lambda \e^{2 \Delta t}}\nonumber\\
 &=& \lim_{t \to \infty} t^{2 \gamma -1}
 \vev{\phi^I (\vec{r} \e^{- \Delta t} \e^{t+\Delta t}) \phi^J
 (\vec{0})}_{\e^{2 \Delta t} m^2 (t+\Delta t, \lambda; g)
 + \Delta t\, C \lambda,\, \lambda
 \e^{2 \Delta t}}\nonumber\\
 &=& \lim_{t \to \infty} t^{2 \gamma -1}
 \vev{\phi^I (\vec{r} \e^{t + \Delta t}) \phi^J (\vec{0})}_{m^2
 (t+\Delta t, \lambda; g),\, \lambda}\nonumber\\
 &=&  \lim_{t \to \infty} t^{2 \gamma -1}
 \vev{\phi^I (\vec{r} \e^{t}) \phi^J (\vec{0})}_{m^2
 (t, \lambda; g),\, \lambda}
\end{eqnarray}
where we have used the RG equation (\ref{linearRG}) going from the
second line to the third.  Thus, we have verified that the right-hand
side of (\ref{relation}) has no dependence on the choice of $\lambda$.
\end{enumerate}

We now rewrite the relation to get an alternative formula along the
line of the strong coupling limit $\lambda_0 \to \infty$ of the
lattice theory.  By integrating the RG equation of the linear sigma
model (\ref{linearRG}), we obtain
\begin{equation}
\vev{\phi^I (\vec{r}) \phi^J (\vec{0})}_{m^2,\, \lambda}
= \vev{\phi^I (\vec{r} \e^{-t}) \phi^J (\vec{0})}_{\e^{2t} 
 (m^2 + C t \lambda), \,  \e^{2 t} \lambda}
\label{linearRGfinite}
\end{equation}
where $t$ is finite.  Hence, we get
\begin{equation}
\vev{\phi^I (\vec{r} \e^t) \phi^J (\vec{0})}_{m^2 (t; \lambda),\,
\lambda}
 = \vev{\phi^I (\vec{r}) \phi^J (\vec{0})}_{\e^{2t}
 (m^2 (t; \lambda) + C t \lambda),\, \e^{2t} \lambda}
\end{equation}
Using (\ref{m2tdep}), we find
\begin{eqnarray}
&&\e^{2t} (m^2 (t, \lambda; g) + C t \lambda) \nonumber\\
 &=& \e^{2t} \left[ (C-1) \lambda  \frac{1}{2} \ln \lambda \e^{2t} + \lambda
 ( \ln \Lambda (g) - c \ln t )  \right]\nonumber\\
 &=& \lambda \e^{2t} \left[ (C-1) \frac{1}{2} \ln \lambda \e^{2t} + 
 \ln \Lambda (g) - c \ln \ln (\lambda \e^{2 t})
 + \mathrm{O} (1/t)  \right]
\end{eqnarray}
Rewriting $\lambda \e^{2t}$ as $\lambda$, the $t \to \infty$ limit on
the right-hand side of (\ref{relation}) can be rewritten as the strong
coupling limit $\lambda \to \infty$:
\begin{equation}
\vev{\Phi^I (\vec{r}) \Phi^J (\vec{0})}_g
 = z \, \left(\frac{g}{1+c g}\right)^{2 \gamma}
 \lim_{\lambda \to \infty} (\ln \lambda)^{2 \gamma - 1}
 \vev{\phi^I (\vec{r}) \phi^J (\vec{0})}_{m^2 (\lambda; g),\, \lambda}
 \label{strong}
\end{equation}
where $m^2 (\lambda; g)$ is given by
\begin{equation}
m^2 (\lambda; g) = \lambda \left[  \frac{C-1}{2} \ln \lambda
 - c \ln \ln \lambda + \ln \Lambda (g) \right]
\label{m2lambda}
\end{equation}
Thus, as expected from the strong coupling limit of the lattice model,
the non-linear sigma model is obtained as an infinite coupling limit
of the linear sigma model.  

\begin{figure}
\includegraphics{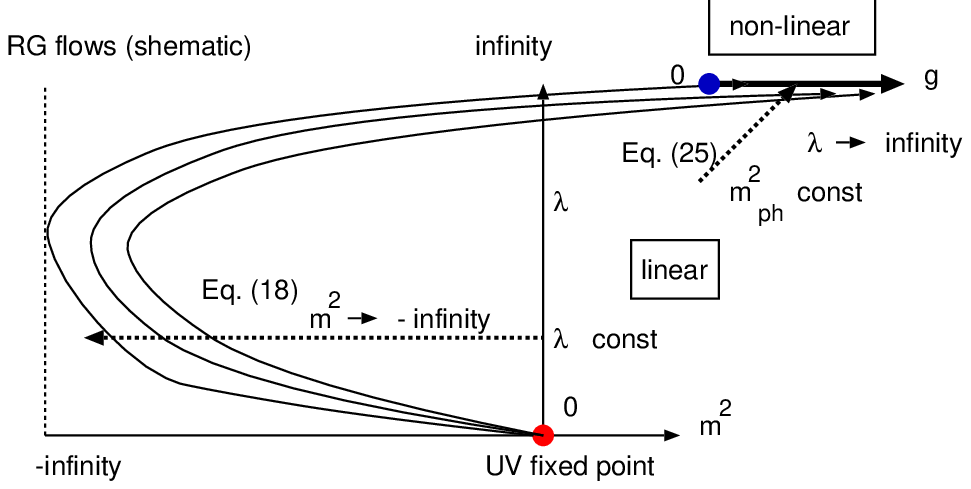}
\caption{Two ways of obtaining the non-linear sigma model from the
linear sigma model.  First, $m^2 \to - \infty$ for a fixed $\lambda$.
Second, $\lambda \to \infty$ with the physical squared mass fixed.  In
the first case, we must further take the infrared limit.}
\end{figure}
We have obtained two relations (\ref{relation}), (\ref{strong}) that
relate the linear sigma model to the non-linear sigma model.  (FIG.~1)
The two are equivalent since they are simply related by the RG
equation (\ref{linearRGfinite}) of the linear sigma model.  To summarize the
main features of the two relations, we find the following:
\begin{itemize}
\item[(\ref{relation})] --- For a fixed $\lambda$, we take $m^2 \to -
\infty$.  The non-linear sigma model is obtained as the infrared limit
of the linear sigma model.
\item[(\ref{strong})] --- Both $\lambda$ and $m^2$ go to infinity in
such a way that the physical mass scale is fixed as $\Lambda (g)$.
\end{itemize}

It is interesting to compare the relation (\ref{strong}) with the
$\lambda_0 \to \infty$ limit of the lattice model.  For the lattice
theory we have found (\ref{latticegzero}), i.e., as $\lambda_0 \to
\infty$ we keep the ratio
\begin{equation}
\frac{- m_0^2}{\lambda_0} = \frac{1}{g_0}
\label{lattice}
\end{equation}
finite.  On the other hand, for (\ref{strong}) we find
\begin{equation}
\frac{- m^2 (\lambda; g)}{\lambda} = - \frac{C-1}{2} \ln \lambda
 + c \ln \ln \lambda + \frac{1}{g} + c \ln \frac{g}{1+c g}
\end{equation}
where we have used (\ref{Lambda}).  Since (\ref{C}) gives
\begin{equation}
C > 1
\end{equation}
we find
\begin{equation}
\frac{ - m^2 (\lambda; g)}{\lambda} 
\stackrel{\lambda \to \infty}{\longrightarrow} - \infty
\end{equation}
Hence, Eq.~(\ref{lattice}), which is valid for the lattice theory,
does not apply to the renormalized theory.  However, if we take the
large $N$ limit, we find $C \to 1$ and $c \to 0$, and we get $\frac{-
m^2}{\lambda} = \frac{1}{g}$ just as in the lattice theory.

To draw some consequences from the relation (\ref{strong}), we examine
the low momentum behavior of the two-point function in both the linear
and non-linear sigma models.  In the non-linear model we expand
\begin{equation}
\int d^2 r\, \e^{i p r} \vev{\Phi^I (\vec{r}) \Phi^J (\vec{0})}_g
 =  \delta^{IJ} 
 \frac{\tilde{z} \cdot \left(\frac{g}{1 + c g}\right)^{2 \gamma}}
 {\mu^2 \Lambda (g)^2 + p^2 + \mathrm{O} (p^4)}
\end{equation}
where $\tilde{z}$ and $\mu$ are constants independent of $g$.
Similarly, in the linear model we expand
\begin{equation}
\int d^2 r \, \e^{i p r} \vev{\phi^I (\vec{r}) \phi^J
 (\vec{0})}_{m^2,\, \lambda}
 =  \delta^{IJ} \frac{Z}{m_{\mathrm{ph}}^2 + p^2
 + \mathrm{O} (p^4)}
\end{equation}
where $Z$ and $m_{\mathrm{ph}}^2$ depend on $m^2$ and $\lambda$.  The
relation (\ref{strong}) implies that if we choose $m^2$ as $m^2
(\lambda; g)$ given by (\ref{m2lambda}), we must find
\begin{eqnarray}
\tilde{z} &=& z\, Z \cdot (\ln \lambda)^{2\gamma - 1}
 \label{norm}\\
\mu^2 \Lambda (g)^2 &=& m^2_{\mathrm{ph}} \label{mass}
\end{eqnarray}
in the limit $\lambda \to \infty$.

We note that the two-point function of the linear model is invariant
under the RG.  Hence, both the normalization constant $Z$ and the
ratio $\frac{m_{\mathrm{ph}^2}}{\lambda}$ are RG invariants.  Hence,
each must be a function of the RG invariant
\begin{equation}
R (m^2,\lambda) \equiv \frac{m^2}{\lambda} - \frac{C}{2} \ln \lambda
\end{equation}
Substituting (\ref{m2lambda}) into the above, we obtain
\begin{equation}
R (m^2 (\lambda; g), \lambda) 
= - \frac{1}{2} \ln \lambda - c \ln \ln \lambda + \ln \Lambda (g) 
\end{equation}
This goes to $- \infty$ as $\lambda \to \infty$.  Hence,
Eqs.~(\ref{norm}, \ref{mass}) imply the following asymptotic behavior:
\begin{eqnarray}
Z (R) &\stackrel{R \to - \infty}{\longrightarrow}& \frac{\tilde{z}}{z}
 \cdot ( - 2 R )^{1 - 2 \gamma} \label{Z} \\
\frac{m^2_{\mathrm{ph}}}{\lambda}
 &\stackrel{R \to - \infty}{\longrightarrow}&
 \mu^2 \left(\e^R (- 2 R)^c \right)^2 \label{ratio}
\end{eqnarray}

The above asymptotic behavior can be checked explicitly in the large
$N$ limit.  In this limit we obtain
\begin{equation}
c = 0,\quad \gamma = \frac{1}{2},\quad C = 1
\end{equation}
Hence, the asymptotic formulas (\ref{Z}, \ref{ratio}) give
\begin{eqnarray}
 Z(R) &\stackrel{R \to - \infty}{\longrightarrow}& \mathrm{const}  
\label{ZN}\\
\frac{m_{\mathrm{ph}}^2}{\lambda} &\stackrel{R \to - \infty}
{\longrightarrow}& \mathrm{const} \cdot \e^{2R} \label{ratioN}
\end{eqnarray}
It is easy to check these:
\begin{enumerate}
\item In the large $N$ limit, the propagator is free:
\begin{equation}
\vev{\widetilde{\phi^I} (p) \phi^J} = \frac{1}{p^2 + m_{\mathrm{ph}}^2}
\end{equation}
This gives $Z = 1$, agreeing with (\ref{ZN}).
\item The physical squared mass $m_{\mathrm{ph}}^2$ is given by
\begin{equation}
m_{\mathrm{ph}}^2 + \frac{\lambda}{2} \ln m_{\mathrm{ph}}^2
 = m^2
\end{equation}
Hence,
\begin{equation}
R \equiv \frac{m^2}{\lambda} - \frac{1}{2} \ln \lambda
 = \frac{m_{\mathrm{ph}}^2}{\lambda}
 + \frac{1}{2} \ln \frac{m_{\mathrm{ph}}^2}{\lambda}
\stackrel{\lambda \to \infty}{\longrightarrow} \frac{1}{2}
 \ln \frac{m_{\mathrm{ph}}^2}{\lambda}
\end{equation}
This implies
\begin{equation}
 \frac{m_{\mathrm{ph}}^2}{\lambda} \stackrel{R \to
 -\infty}{\longrightarrow} \e^{2 R}
\end{equation}
agreeing with (\ref{ratioN}).
\end{enumerate}
In the large $N$ limit, we can also check easily that the four-point
function (in momentum space) of the linear sigma model reduces to that
of the non-linear sigma model in the strong coupling limit if we fix
the physical squared mass (FIG.~2):
\begin{figure}
\includegraphics{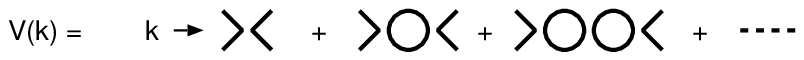}
\caption{Four-point vertex in the momentum space in the large $N$
limit}
\end{figure}
\begin{eqnarray}
 V (k) &=& - \frac{1}{N} \cdot \frac{1}{ 
\frac{1}{4 \pi \lambda} +
\frac{1}{2\pi} 
\frac{1}{\sqrt{k^2(k^2+ 4 m^2_{\mathrm{ph}})}} ~\mathrm{arctanh}~
\sqrt{\frac{k^2}{k^2+ 4 m^2_{\mathrm{ph}}}} }\nonumber\\
 &\stackrel{\lambda \to \infty}{\longrightarrow}&
- \frac{1}{N} \cdot \frac{1}{ 
\frac{1}{2\pi} 
\frac{1}{\sqrt{k^2(k^2+ 4 m^2_{\mathrm{ph}})}} ~\mathrm{arctanh}~
\sqrt{\frac{k^2}{k^2+ 4 m^2_{\mathrm{ph}}}} }
\end{eqnarray}

In conclusion we have derived two formulas (\ref{relation},
\ref{strong}) giving the two dimensional $O(N)$ non-linear sigma model
as a limit of the linear sigma model.  Especially as a consequence of
(\ref{strong}) we have obtained the asymptotic behavior (\ref{Z},
\ref{ratio}).  The relation discussed in this paper is not limited to
the $O(N)$ sigma models, and it is an example of a general relation
between the linear and non-linear models in two and three dimensions.
We summarize the analogous results for the Gross-Neveu model
(non-linear) and the Yukawa model (linear) in two dimensions in
appendix B.

\appendix

\section{The $O(N)$ sigma models in $3$ dimensions}

For three dimensions the relation between the two types of sigma
models has been well understood for a long time.\cite{wk74,parisi} We
wish to briefly review this case, since the results we obtain for the
two dimensional models are similar to those for the three dimensional
models.

The $O(N)$ linear sigma model, defined in three dimensional euclidean
space, is a superrenormalizable theory.  It is parametrized by a
squared mass $m^2$ and a self-coupling constant $\lambda$ satisfying
the renormalization group (RG) equations:
\begin{eqnarray}
\frac{dm^2}{dt} &=& 2 m^2 + C \lambda^2\\
\frac{d\lambda}{dt} &=& \lambda
\end{eqnarray}
where $\lambda$ is normalized so that 
\begin{equation}
C = - \left(\frac{N}{2} + 1 \right) \frac{1}{(4\pi)^2} < 0
\end{equation}
Under the renormalization group, 
\begin{equation}
R \equiv \frac{m^2}{\lambda^2}  - C \ln \lambda
\end{equation}
is invariant.  For a given $\lambda$, there is a value $m^2_{cr}
(\lambda)$ of the squared mass at which the theory becomes critical.
If we let $R_{cr}$ be the value of $R$ for the critical theory, then
we obtain
\begin{equation}
m^2_{cr} (\lambda) = \lambda^2 ( R_{cr} + C \ln \lambda )
\end{equation}
The scalar field has no anomalous dimension, and the two-point
function satisfies the RG equation
\begin{equation}
\vev{\phi^I (\vec{r} \e^{- \Delta t}) \phi^J (\vec{0})}_{\e^{2 \Delta
t} (m^2 + \Delta t \, C \lambda^2),\, \e^{\Delta t} \lambda}
 = \e^{\Delta t} \vev{\phi^I (\vec{r}) \phi^J (\vec{0})}_{m^2,\,
 \lambda} \label{linearRG3}
\end{equation}

The renormalization of the three-dimensional $O(N)$ non-linear sigma
model is less well known, and we start from a theory defined on a
cubic lattice.  The action is given by
\begin{equation}
S \equiv - \frac{1}{g_0} \sum_{\vec{n}} \sum_{i=1}^3 \left(
\Phi^I_{\vec{n}+\hat{i}} - \Phi^I_{\vec{n}}\right)^2
\end{equation}
Let $g_{0,cr}$ be the critical value.  For $g_0 > g_{0,cr}$, the
theory is $O(N)$ symmetric, but for $g_0 < g_{0,cr}$ the symmetry is
spontaneously broken to $O(N-1)$.  The critical point is characterized
by two critical indices:
\begin{enumerate}
\item $y_E$ --- Near criticality, the correlation length $\xi$ (or inverse
physical mass) behaves as
\begin{equation}
\xi \propto \left| g_0 -  g_{0,cr} \right|^{\frac{1}{y_E}}
\end{equation}
\item $\eta$ --- Near but below criticality, the VEV of $\Phi^I$ behaves as
\begin{equation}
\vev{\Phi^I} \propto (g_{0,cr} - g_0)^{\frac{1 + \eta}{2}}
\end{equation}
\end{enumerate}
The critical indices can be calculated in various ways.  For example,
to lowest order in the $\epsilon$ expansion, we find \cite{wk74}
\begin{eqnarray}
y_E &=& 2 - \frac{N+2}{N+8} \,\ep\\
 \eta &=& \frac{N+2}{2(N+8)^2}\, \ep^2
\end{eqnarray}
where $\ep = 1$ for three dimensional space.  Using the critical
indices, the continuum limit of the two-point function is defined by
\begin{equation}
\vev{\Phi^I (\vec{r}) \Phi^J (\vec{0})}_{g}
 \equiv \lim_{t \to \infty} \e^{t(1+\eta)}
 \vev{\Phi^I_{\vec{n}=\vec{r}\e^t} \Phi^J_{\vec{0}}}_{g_0
 = g_{0,cr} + g \e^{- y_E t}}
\end{equation}
This satisfies a simple RG equation
\begin{equation}
\vev{\Phi^I (\vec{r} \e^{- \Delta t}) \Phi^J (\vec{0})}_{g \e^{y_E
\Delta t}} = \e^{\Delta t \, (1+\eta)}
 \vev{\Phi^I (\vec{r}) \Phi^J (\vec{0})}_g
\end{equation}

We have two ways of obtaining the non-linear sigma model as a limit of
the linear sigma model. (FIG.~3) One is analogous to (\ref{relation}),
and the other to (\ref{strong}).\footnote{Actually, it is more
appropriate to say that (\ref{relation}) is analogous to the well
known (\ref{relation3}), and (\ref{strong}) to (\ref{strong3}).}
First, the analog of (\ref{relation}) is given by
\begin{equation}
\vev{\Phi^I (\vec{r}) \Phi^J (\vec{0})}_g
 = z \lambda^\eta \lim_{t \to \infty} \e^{(1+\eta)t}
 \vev{\phi^I (\vec{r}\e^t) \phi^J (\vec{0})}_{m^2 = m^2_{cr} (\lambda)
 + z_m \lambda^{2-y_E} g \e^{- y_E t},\, \lambda}
 \label{relation3}
\end{equation}
where $z,\, z_m$ are numerical constants.  The analog of
(\ref{strong}) is obtained from the above by using the RG equation
(\ref{linearRG3}) as
\begin{equation}
\vev{\Phi^I (\vec{r}) \Phi^J (\vec{0})}_g
 = z \lim_{\lambda \to \infty} \lambda^\eta \vev{\phi^J (\vec{r})
 \phi^J (\vec{0})}_{m^2 = m_{cr}^2 (\lambda) + z_m \lambda^{2-y_E}
 g,\, \lambda} \label{strong3}
\end{equation}
\begin{figure}
\includegraphics{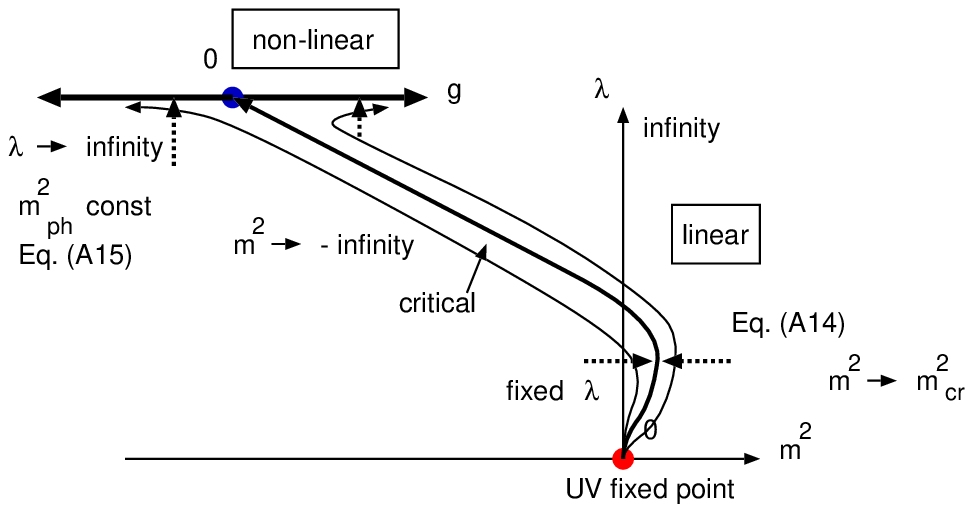}
\caption{Two ways of obtaining the non-linear sigma model from the
linear sigma model in three dimensions.}
\end{figure}

Let us introduce the low momentum expansion of the Fourier transform
of the two-point functions.  For simplicity we restrict ourselves to
the symmetric phase.  Then, we obtain
\begin{eqnarray}
\int d^3 r\, \e^{- i p r} \vev{\phi^I (\vec{r}) \phi^J
(\vec{0})}_{m^2,\, \lambda} &=& \delta^{IJ} 
 \frac{\tilde{z} (R)}{m_{\mathrm{ph}}^2 + p^2 + \cdots}\\
\int d^3 r\, \e^{- i p r} \vev{\Phi^I (\vec{r}) \Phi^J (\vec{0})}_g
 &=& \delta^{IJ} \frac{Z \cdot g^{\frac{\eta}{y_E}}}
 {\mu^2 g^{\frac{2}{y_E}} + p^2 + \cdots}
\end{eqnarray}
where $Z,\, \mu$ are constants.  The relation (\ref{strong3}) implies
the following asymptotic behavior\cite{parisi}
\begin{eqnarray}
\tilde{z} (R) &\stackrel{R \to R_{cr}}{\longrightarrow}&
 \frac{Z}{z} \left(\frac{R-R_{cr}}{z_m}\right)^{\frac{\eta}{y_E}}\\
\frac{m_{\mathrm{ph}}^2}{\lambda^2} &\stackrel{R \to
 R_{cr}}{\longrightarrow}&
 \mu^2 \left(\frac{R-R_{cr}}{z_m}\right)^{\frac{2}{y_E}}
\end{eqnarray}

\section{The Gross-Neveu model vs. the Yukawa model}

In this appendix we briefly discuss how to obtain the Gross-Neveu
model in two dimensions\cite{gn}
\begin{equation}
\mathcal{L}_{GN} = \bar{\psi}^I \frac{1}{i} \fmslash{\partial} \psi^I
 + \frac{g}{2 N} \left( \bar{\psi}^I \psi^I \right)^2
\end{equation}
(where $I=1,\cdots,N$) as a strong coupling limit of the Yukawa model
\begin{equation}
\mathcal{L}_Y = \bar{\psi}^I \frac{1}{i} \fmslash{\partial} \psi^I
+ \frac{1}{2} (\partial_\mu \phi)^2 + \frac{M^2}{2} \phi^2 +
i \frac{y}{\sqrt{N}} \phi \, \bar{\psi}^I \psi^I
\end{equation}

Both theories are invariant under the $\mathbf{Z}_2$ transformation:
\[
\psi^I \to \gamma_5 \psi^I,\quad
\bar{\psi}^I \to - \bar{\psi}^I \gamma_5,\quad
(\phi \to - \phi\quad\textrm{for the Yukawa model})
\]
In the following we will only consider the case
\[
N > 1
\]
Note that, using the $\mathbf{Z}_2$ transformation,
we can adopt the convention
\[
y > 0
\]

We first recall the RG equations of the renormalized parameters.  For
the Gross-Neveu model we have\cite{wetzel}
\begin{equation}
\frac{d}{dt} g = \beta_1 g^2 + \beta_2 g^3
\end{equation}
where
\begin{equation}
\beta_1 = \frac{1}{\pi}\frac{N-1}{N},\quad \beta_2 = - \frac{1}{2
\pi^2}\frac{N-1}{N^2}
\end{equation}
By rewriting $\beta_1 g$ as $g$, we can rewrite the RG equation as
\begin{equation}
\frac{d}{dt} g = g^2 + c g^3
\end{equation}
where
\begin{equation}
c \equiv \frac{\beta_2}{\beta_1^2}
= - \frac{1}{2(N-1)}
\end{equation}
On the other hand, for the Yukawa model we have
\begin{equation}
\left\lbrace\begin{array}{c@{~=~}l}
\frac{d M^2}{dt} & 2 M^2 + C y^2\\
\frac{d y}{dt} & y
\end{array}\right.
\end{equation}
where
\begin{equation}
C = - \frac{1}{\pi} 
\end{equation}
An RG invariant is obtained as
\begin{equation}
R (M^2, y) \equiv \frac{M^2}{y^2} - C \ln y
\end{equation}

The physics of the models can be summarized as follows:
\begin{enumerate}
\item The $\mathbf{Z}_2$ symmetry is spontaneously broken in both
theories.  For the Yukawa model, the symmetry is broken irrespective
of the choice of $M^2$ and $y$ as long as $y \ne 0$.
\item In the Gross-Neveu model, the mass of the fermions is a constant
multiple of the mass scale
\begin{equation}
\Lambda (g) \equiv \e^{- \frac{1}{g}} \left(\frac{g}{1 + c
g}\right)^{-c}
\end{equation}
\item In the Yukawa model, the mass of the fermions can be expressed
as
\begin{equation}
m_{\mathrm{ph}} = y f(R)
\end{equation}
where $f(R)$ is an unknown RG invariant function.  We will obtain the
asymptotic behavior of $f(R)$ for $R \to \infty$ at the end.
\end{enumerate}

We proceed in the same way as for the two dimensional sigma models
discussed in the main text.  We first obtain a guess for the relation
between the two models by a naive manipulation of the lagrangian.
Then, we improve the relation using the renormalization group as the
guiding principle.  The asymptotic freedom of the Gross-Neveu model is
the key ingredient of the derivation.

For a finite $y$, let us take $M^2 > 0$ very large so that the kinetic
term of the scalar field can be ignored in comparison to the potential
term.  We rewrite the lagrangian as
\begin{eqnarray}
\mathcal{L}_Y &=& \bar{\psi}^I \frac{1}{i} \fmslash{\partial} \psi^I +
\frac{1}{2} \frac{y^2}{M^2} \frac{1}{N} \left(\bar{\psi}^I \psi^I
\right)^2 + \frac{1}{2} (\partial_\mu \phi)^2\nonumber\\
&& + \frac{1}{2} M^2 \left( \phi + \frac{y}{\sqrt{N} M^2} i
\bar{\psi}^I \psi^I \right)^2
\end{eqnarray}
In the limit $M^2 \to + \infty$ we get the constraint
\begin{equation}
\phi = - \frac{y}{\sqrt{N} M^2} i \bar{\psi}^I \psi^I
\end{equation}
and the lagrangian reduces to that of the Gross-Neveu model with a
small coupling $g_0$ where
\begin{equation}
\frac{g_0}{\beta_1} \equiv \frac{y^2}{M^2} \ll 1
\end{equation}
Since the Gross-Neveu model is asymptotic free, we can regard $g_0$ as
the coupling at short distances.  Recalling the dependence of the
running coupling on the logarithmic distance scale $t \gg 1$
\begin{equation}
g(-t) \simeq \frac{1}{t + c \ln t - \ln \Lambda (g)} \ll 1
\end{equation}
we find that the necessary $t$-dependence of $M^2$ is given by
\begin{equation}
M^2 = y^2 \beta_1 \left( t + c \ln t - \ln \Lambda (g) +
\left( 1 + \frac{C}{\beta_1} \right) \ln y \right)
\end{equation}
where the $y$-dependent finite term is added so that the limit we are
about to write down does not depend on the choice of $y$.

Using the result of the naive manipulation as a hint, we guess that
the Gross-Neveu model is obtained as the following limit of the Yukawa
model:
\begin{equation}
\vev{\psi^I (\vec{r}) \bar{\psi}^J (\vec{0})}_g = z \lim_{t \to
\infty} \e^t \vev{\psi^I (\vec{r} \e^t) \bar{\psi}^J (\vec{0})}_{M^2
= y^2 \beta_1 \left( t + c \ln t - \ln \Lambda (g) +
\left( 1 + \frac{C}{\beta_1} \right) \ln y \right) \atop y \hfill}
\end{equation}
where $z$ is an unknown finite constant of order $1$.  Using the RG
equation of the Yukawa model
\begin{equation}
\vev{\psi^I (\vec{r} \e^{\Delta t}) \bar{\psi}^J (\vec{0})}_{M^2, y}
= \e^{- \Delta t} \vev{\psi^I (\vec{r}) \bar{\psi}^J (\vec{0})}_{\e^{2
\Delta t} (M^2 + C \Delta t\, y^2),\, \e^{\Delta t} y}
\end{equation}
we can check that the limit does not depend on the choice of $y$.
Using the RG equation further, we can rewrite the above limit as the
strong coupling limit:
\begin{equation}
\vev{\psi^I (\vec{r}) \bar{\psi}^J (\vec{0})}_g
= z \lim_{y \to \infty} \vev{\psi^I (\vec{r}) \bar{\psi}^J
(\vec{0})}_{M^2 = y^2 \beta_1 \left( \left(1 + \frac{C}{\beta_1}
\right) \ln y + c \ln \ln y - \ln \Lambda (g) \right)\atop
y\hfill}
\end{equation}
Since 
\begin{equation}
1 + \frac{C}{\beta_1} = - \frac{1}{N-1} < 0
\end{equation}
$M^2 \to - \infty$ in the strong coupling limit.  We also find
the following asymptotic behavior of the RG invariant
\begin{equation}
R \equiv \frac{M^2}{y} - C \ln y = \beta_1 \left(
\ln y + c \ln \ln y - \ln \Lambda (g) \right) \stackrel{y \to +
\infty}{\longrightarrow} + \infty \label{M2}
\end{equation}
This relation between $y$ and $R$ guarantees the finiteness of the
physical fermion mass $m_{\mathrm{ph}}$ as we take $y \to + \infty$.

Eq.~(\ref{M2}) implies
\begin{equation}
\Lambda (g) = y \cdot \left( \ln y \right)^c \e^{- \frac{R}{\beta_1}}
\stackrel{y \to + \infty}{\longrightarrow} \frac{1}{\beta_1^c} y R^c
\, \e^{- \frac{R}{\beta_1}}
\end{equation}
Since the fermion mass is a constant multiple of $\Lambda (g)$, we
obtain the following asymptotic behavior for the Yukawa model:
\begin{equation}
\frac{m_{\mathrm{ph}}}{y} \equiv f(R) \stackrel{R \to +
\infty}{\longrightarrow} \mathrm{const}\cdot R^c \e^{-
\frac{R}{\beta_1}} \label{mph}
\end{equation}

In the large $N$ limit, we find
\begin{equation}
\beta_1 = - C = \frac{1}{\pi},\quad c = 0
\end{equation}
and 
\begin{equation}
\left\lbrace\begin{array}{c@{~=~}l}
 M^2 & \frac{1}{\pi} y^2 \frac{1}{g}\\
 \frac{m_{\mathrm{ph}}}{y} & \e^{- \pi R}
\end{array}\right.
\end{equation}
This is consistent with Eqs.~(\ref{M2}, \ref{mph}).

\begin{acknowledgments}
The original version of the paper was written during a visit to UCLA.
I would like to thank Prof. T.~Tomboulis for hospitality.  This work
was partially supported by the Grant-In-Aid for Scientific Research
from the Ministry of Education, Culture, Sports, Science, and
Technology, Japan (\#14340077).
\end{acknowledgments}

\bibliography{sigma2d}

\end{document}